\documentclass[aps,prl,twocolumn,groupedaddress,showpacs,amsmath]{revtex4}
\usepackage{graphicx, epsfig, textcomp, wasysym, hyperref}
\begin{document}

\preprint{APS/123-QED}

\title{Periodicity in Al/Ti superconducting single electron transistors}%

\author{Sarah J. MacLeod, Sergey Kafanov and Jukka P. Pekola}
\affiliation{Low Temperature Laboratory, Helsinki University of Technology, P.O. Box 3500, FIN-02015 TKK, Finland}
\date{\today}

\begin{abstract}
We present experiments on single Cooper-pair transistors made of two different superconducting materials. We chose Ti and Al to create an energy gap profile such that the island has a higher gap than the leads, thereby acting as a barrier to quasiparticle tunneling. Our transport measurements demonstrate that quasiparticle poisoning is suppressed in all our TiAlTi structures (higher gap for the island) with clear $2e$ periodicity observed, whereas full quasiparticle poisoning is observed in all AlTiAl devices (higher gap for the leads) with $e$ periodicity.
\end{abstract}

\maketitle
Superconducting single electron transistors (SSETs) consist of a superconducting island weakly coupled to two superconducting leads via low capacitance, high resistance Josephson junctions and a capacitive gate. The gate charge $Q_g=C_g V_g$, given by the gate capacitance $C_g$ and applied voltage $V_g$, controls the number of electrons on the island. Thus current through the SSET is expected to peak every $2e$ in gate charge for bias voltages at which the Coulomb energy barrier for Cooper pairs is removed. This $2e$-periodicity is a signature of current carried by Cooper pairs and not by single electrons [quasiparticles (QPs)]. However though $2e$ periodicity is expected, $e$ periodic behavior can be observed \cite{Eiles,Wal}. This is believed to be a consequence of non-equilibrium QPs, which exist even at low temperatures where thermal QPs would ideally vanish. One parameter which has been shown to affect QP tunneling on and off the SSET island is the relative difference between superconducting gaps of the leads and the island \cite{Chi}. If the island gap is made larger than the lead gap this will increase the QP energy on the island, which in turn favors $2e$-periodicity at low temperatures. Previous studies have used Al-AlO$_{x}$-Al SSETs, where the difference in the gaps of the leads and the island has been achieved by flowing oxygen during the deposition of one of the Al layers \cite{Aumen} or by varying the thickness of the Al layer \cite{Yam,court,court2,ferg}.

\begin{figure}
\centering\epsfig{figure=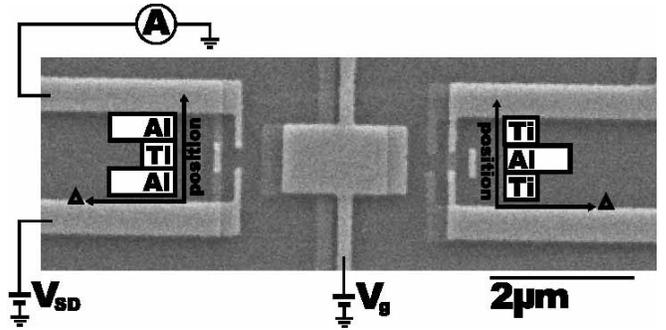,width=\linewidth}
\caption{{\label{fig:1}} Scanning electron micrograph of the mirror pair device (b), with the measurement set-up for the AlTiAl device drawn. Inset: a schematic of the gap profiles for the AlTiAl SSET (left) and TiAlTi SSET (right).}
\end{figure}

An alternative to these works is to use two materials with naturally different superconducting gaps, such as aluminium and niobium. However such junctions suffered from large leakage current and QP poisoning \cite{Nb1,Nb2,Nb3,Nb4,Nb5,Nb6,Nb7}. Previous research has also used Ta and Al to trap QPs in X-ray detection applications \cite{Gaidis} and to study thermodynamic fluctuations \cite{Wilson}. In the same spirit, here we have fabricated and measured the first SSETs from Al and Ti. These materials have a large difference in gaps ($\Delta_\mathrm{Ti}\simeq 200~\mathrm{\mu eV}$, $\Delta_\mathrm{Al}\simeq 50~\mathrm{\mu eV}$). We fabricated mirror devices on the same chip, comprising of AlTiAl and TiAlTi SSETs. Figure\,\ref{fig:1} shows a scanning electron micrograph of the mirror pair SSET device (b). The brighter areas are the evaporated Ti. The device on the right is a TiAlTi SSET and on the left is an AlTiAl SSET with the measurement setup illustrated. Additionally, schematics of the gap profiles are shown inset. As Ti has a smaller gap than Al, the Ti island acts as a QP well in the AlTiAl device, hence this SSET should exhibit $e$ periodicity. In contrast $2e$ periodicity is expected in the TiAlTi SSET as it is energetically unfavorable for a QP to tunnel onto and stay on the Al island.
\begin{table*}
\caption{Sample parameters of AlTiAl and TiAlTi SSETs. Here the sum of the Al and Ti gaps is given by $\Delta_\mathrm{Al}+\Delta_\mathrm{Ti}$, $R_N$ is the normal-state resistance of the two junctions in serie-, $E_c$ is the charging energy, $E_\mathrm{J}$ is the Josephson energy, $D_i$ corresponds to the $i\,\mathrm{th}$ deposition, Ox gives the oxidation pressure and time and $P$ is the observed periodicity of the gate modulations.}
\begin{ruledtabular}
\begin{tabular}{ccccccccc}
Sample & $\Delta_\mathrm{Al}+\Delta_\mathrm{Ti}\,(\mathrm{\mu eV})$ & $R_N\,(\mathrm{k\Omega})$ & $E_\mathrm{J}\,(\mu\mathrm{eV})$ & $E_c\,(\mu\mathrm{eV})$ & D1 (nm) & Ox (mbar/min) & D2 (nm) & P\\
\hline
$\mathrm{AlTiAl}_{a}$ & 280 & 25.3 & 25.7 & 255 & Al (20)& $0.24/15$ & Ti (40) & $e$ \\       % 14 Jan 2009
$\mathrm{TiAlTi}_{a}$ & 280 & 27.6 & 23.6 & 255 & Al (20)& $0.24/15$ & Ti (40) & $2e$ \\       % 14 Jan 2009
$\mathrm{AlTiAl}_{b}$ & 250 & 12.7 & 45.7 & 250 & Al (20) & $0.11/10$ & Ti (30) & $e$ \\     % 18 Dec 2009
$\mathrm{TiAlTi}_{b}$ & 250 & 13.8 & 42.1 & 210 & Al (20) & $0.11/10$ & Ti (30) & $2e$ \\     % 18 Dec 2009
$\mathrm{AlTiAl}_{c}$ & 220 & 9.2 & 55.6 & 195 & Al (20) & $0.26/15$ & Ti (40) & $e$ \\     % 21 Jan 2009
$\mathrm{TiAlTi}_{c}$ & 240 & 12.7 & 43.9 & 250 & Al (20) & $0.26/15$ & Ti (40) & $2e$ \\     % 21 Jan 2009
\end{tabular}\label{tab:table1}
\end{ruledtabular}
\end{table*}
The samples were fabricated by electron-beam lithography using the standard shadow deposition technique \cite{Dolan}. Each mirror pair device was fabricated in one vacuum cycle. Measurements were made at temperatures from $50\,\mathrm{mK}$ up to about $300\,\mathrm{mK}$. The source-drain current is measured as a function of voltage bias, $V_b$ and gate voltage, $V_g$ which is plotted as a stability diagram for the three AlTiAl and TiAlTi SSET mirror pairs measured. The stability diagrams were measured using a two-probe, voltage-biased configuration. From these measurements we extracted the sample parameters as detailed in Table\,\ref{tab:table1}, with pairs being denoted by the same subscript. Our  samples typically had a series resistance of $R_N\sim 10-20\,\mathrm{k\Omega}$ across the junctions and with Josephson energy, $E_\mathrm{J} \sim 30-50\,\mathrm{\mu eV}$. $E_\mathrm{J}$ was extracted using $E_\mathrm{J}=(\hbar/2e)I_c$, where $I_c$ is the critical current given by \cite{barone}
\begin{equation}
I_c=\frac{2}{e(R_N/2)}\frac{\Delta_\mathrm{Ti}\Delta_\mathrm{Al}} {\Delta_\mathrm{Ti}+\Delta_\mathrm{Al}}K\left(\left|\frac{\Delta_\mathrm{Al}- \Delta_\mathrm{Ti}}{\Delta_\mathrm{Ti}+\Delta_\mathrm{Al}}\right|\right),
\end{equation}
where $R_N$ is the normal-state in series resistance given in Table\,\ref{tab:table1}, $K(x)$ is an elliptic integral and we assume $\Delta_\mathrm{Al}\sim 5\Delta_\mathrm{Ti}$ (see discussion which follows). The charging energy is roughly $E_c\sim 200\,\mathrm{\mu eV}$. We extracted $E_c$ as well as the sum of the Ti and Al gaps, $\Delta_\mathrm{Ti}+\Delta_\mathrm{Al}$, from the stability diagrams by considering the energy requirements for different tunneling processes through the island. For a given tunneling event, the bias voltage must supply sufficient energy in order to overcome the change in the charging energy $\Delta U=U(n\pm m)-U(n)$ caused by transferring $m$ electrons on/off the island. Here the charging energy $U(n)=(Q_{g}-ne)^{2}/2C_{\Sigma}$ is determined by the number of excess electrons on the island, $n$, induced gate charge, $Q_{g}$, and the sum of the capacitances, $C_{\Sigma}=C_1+C_2+C_g$, where $C_i$ is the capacitance of the $i^{\mathrm{th}}$ junction and $C_g$ is the capacitance of the gate.

For $V_b>0$ the condition at the $i^{\mathrm{th}}$ junction to transfer a charge of $m$ and with $q$ quasiparticles created is given by \cite{fitz}
\begin{equation}
\begin{split}
m\kappa_{i}eV_{b}&=U(n\pm{}m)-U(n)+q(\Delta_\mathrm{Ti}+\Delta_\mathrm{Al})\\
&=2mE_c\left[(-1)^{i-1}\left(\frac{Q_{g}}{e}-n\right)+\frac{m}{2}\right]\\
&+q(\Delta_\mathrm{Ti}+\Delta_\mathrm{Al}).
\end{split}
\label{eqn:1}
\end{equation}
Here $\kappa_{i}=1-(C_i+C_g/2)/C_{\Sigma}$ is the fraction of the bias voltage across junction $i$ for $i=1,2$.
\begin{figure}
\centering\epsfig{figure=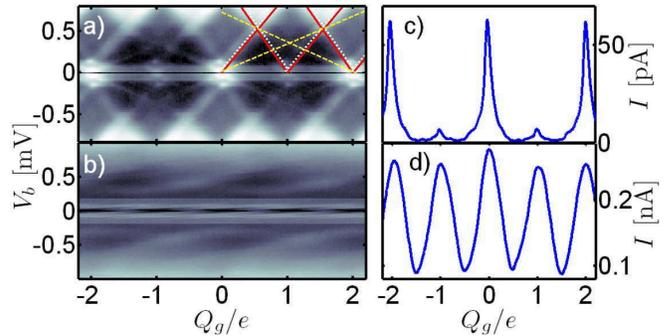,width=\linewidth}
\caption{{\label{fig:2}}(a) Stability diagram of $\mathrm{TiAlTi}_{a}$ on the plane of bias $V_b$, and normalized gate voltage, $Q_g/e$. To emphasize the observed current features we plot $\log|I|$; the white (dark) areas correspond to higher (lower) current with a range of $|I|=100\,\mathrm{pA}$. We modeled the voltage conditions for Cooper pair tunneling (solid red line), QP tunneling (dotted white line) and $3e$ tunneling (dashed yellow line) (b) Stability diagram of $\mathrm{AlTiAl}_c$ with $I$ given in arbitrary units, again lighter areas correspond to higher current. (c) Cross-section of Fig.\,\ref{fig:2}(a) at constant bias voltage, $V_b=0.06\,\mathrm{mV}$. (d) Cross-section of Fig.\,\ref{fig:2}(b) at constant bias voltage, $V_b=0.06\,\mathrm{mV}$.}
\end{figure}

Figure\,\ref{fig:2}(a) shows the condition for resonant Cooper pair tunneling given by the solid red line ($m=2$, $q=0$) as well as the threshold above which QP tunneling is allowed given by the dotted white line ($m=1$, $q=1$). Here lines with a positive (negative) slope are for tunneling processes which transfer a charge $m$ off (on) the island through junction 1 (2). The bright spots around $V_b=0$, where the solid red lines cross, are supercurrent features due to Cooper pair tunneling. The Josephson quasiparticle cycles are the diamond-like structures occurring for $|V_b|>0.5\,\mathrm{mV}$, along the resonant Cooper pair tunneling condition (solid red line) and above the quasiparticle threshold (dotted white lines). The three points of the ``saw-tooth'' feature at odd values of $V_g$ are due to $3e$ or a combination of $3e$ and QP tunneling processes. $3e$ tunneling is a higher-order process due to a Cooper pair tunneling on (off) the island while simultaneously a QP tunnels off (on), thereby effectively transferring a charge of $e$ to (from) the island. The central apex of the saw-tooth feature coincides with the intersection of opposite $3e$ tunneling conditions (dashed yellow lines), hence this feature is believed to be due to a charge of $e$ tunneling through the island by $3e$ processes. This is favorable when $V_{b}$ satisfies \cite{pohlen}

\begin{figure}
\centering\epsfig{figure=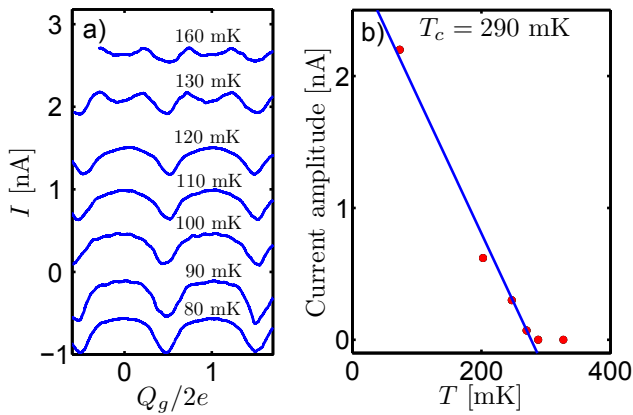,width=\linewidth}
\caption{{\label{fig:4}}(a) Gate modulations in $I$ versus gate voltage for $\mathrm{TiAlTi}_{c}$ with increasing temperature. Curves are vertically offset by $0.5\,\mathrm{nA}$ for clarity. Each temperature measurement was performed separately, therefore curves were horizontally aligned to account for charge jumps. (b) Amplitude of the gate modulation in $I$ versus temperature for a TiAlTi SSET fabricated to extract the $T_c$ of Ti.}
\end{figure}
\begin{equation}
\begin{split}
(m_{1}\kappa_{1}+m_2\kappa_2)eV_{b}&=U(n\pm\delta m)-U(n)\\
&+q(\Delta_\mathrm{Ti}+\Delta_\mathrm{Al}).
\end{split}
\label{eqn:2}
\end{equation}
Here $\delta m=m_2-m_1$ gives the charge transferred through the island by $m_1$ electrons tunneling off via junction 1 as $m_2$ electrons simultaneously tunnel on via junction 2. The two points on either side of the central apex occur along the $3e$ tunneling condition (dashed yellow line) and above the QP threshold (white dotted line). The right (left)-most feature (at $V_b>0$) is due to a charge of $e$ tunneling off (onto) the island via $3e$ tunneling followed by a QP tunneling event on (off) the island.

The TiAlTi SSET sample has a higher gap on the island (Fig.\,\ref{fig:2}(a)), whereas the AlTiAl SSET has a higher gap for the leads, as shown in Fig.\,\ref{fig:2}(b). Figure\,\ref{fig:2}(a) has a clear $2e$ periodic structure for the TiAlTi SSET in the current around $V_b=0$, with slight quasiparticle poisoning (the spots occurring at $V_g=(2n+1)Q_g/e$ for integer $n$). This is highlighted in Fig.\,\ref{fig:2}(c), which shows a cross-section of Fig.\,\ref{fig:2}(a) at constant $V_b$. Here small peaks in the current occur at odd values of $V_g$ in addition to the main peaks at even values of $V_g$. In the AlTiAl SSETs (Fig.\,\ref{fig:2}(b)) only $e$ periodic gate modulations are observed. This is emphasized in Fig.\,\ref{fig:2}(d), which shows $e$-periodic modulations with $V_g$ at constant $V_b$ .

Gate modulations of sample $\mathrm{TiAlTi}_{c}$ at different bath temperatures are shown in Fig.\,\ref{fig:4}(a), with the transition from $2e$ to $e$ with increasing temperature due to an increase in the concentration of thermal QPs \cite{Tuom}. In Fig.\,\ref{fig:4}(b) we estimate the energy gaps of Ti and Al to differ by a factor of $\sim 5$ by measuring a TiAlTi SSET which was not fabricated into a mirror pair. It had an in series normal resistance of $R_N=52\,\mathrm{k\Omega}$, periodicity, $P=2e$, and an Al island of thickness $30\,\mathrm{nm}$ and Ti leads of thickness $25\,\mathrm{nm}$. The bath was heated until the gate modulations were suppressed. We plot the amplitude of the gate modulations in $I$ versus temperature at a small constant bias voltage. We fit a straight line to the data and the intersection point where the gate modulations tend to zero gives an estimate of the critical temperature of Ti, $T_{c}^\mathrm{Ti}\simeq 290\,\mathrm{mK}$. Using the relation $\Delta(0)\simeq 1.76k_{\mathrm{B}}T_{c}$ \cite{Tink} we extract $\Delta_\mathrm{Ti}=41\,\mathrm{\mu eV}$ at $T=0\,\mathrm{K}$. This indicates the gap of Ti is roughly 5 times smaller than that of Al.

In summary, we have exploited the large difference between the superconducting gaps of Ti and Al to suppress QP poisoning in SSETs. We have measured the gate modulation of our SSETs, in which $2e$ periodicity was observed in all the samples where the gap was larger on the island than in the leads. In contrast, $e$ periodicity was observed in all the SSETs with an opposite gap profile. Our observations demonstrate that Ti and Al can be used to control the gap profile of SSETs to suppress QP poisoning.

We thank Tommy Holmqvist for assistance in sample fabrication and Yuri Pashkin for useful discussions. This work is partially supported by the European Community's Seventh Framework Program under Grant Agreement No.218783 (SCOPE).

\end{document}